\documentclass[a4paper,11pt]{article}
\usepackage{amsmath,amssymb,amsthm}

\usepackage[pdfpagemode=UseOutlines,
 bookmarks=true,bookmarksopen=true,bookmarksnumbered=true,
 pdffitwindow=true,pdfpagelayout=SinglePage,pdfhighlight=/P,
 colorlinks=true,linkcolor=blue,citecolor=blue,urlcolor=blue]{hyperref}

\theoremstyle{plain}
\newtheorem{theorem}{Theorem}
\newtheorem{statement}[theorem]{Statement}

\title{On a discrete analog of the Tzitzeica equation}
\author{V.E. Adler\\[1ex]
\em\small L.D.~Landau Institute for Theoretical Physics,\\
\em\small 1A Ak.~Semenov, 142432 Chernogolovka, Russia.\\
\em\small E-mail: adler@itp.ac.ru}
\date{March 26, 2011}

\begin{document}\maketitle

\noindent\hrulefill\par
\noindent{\bf Abstract.} A discrete analog of the Tzitzeica equation is
found in the form of quad-equation. Its continuous symmetry is an inhomogeneous
Narita--Bogoyavlensky type lattice equation which defines a discretization of
the Sawada--Kotera equation. The integrability of these discretizations is
proven by construction of the Lax representations.\medskip

\noindent Keywords: Tzitzeica equation, discrete differential geometry, Lax pair,
higher symmetry
\par
\noindent\hrulefill

%-------------------------------------------------------------------------------
\section{Introduction}

The celebrated Tzitzeica equation \cite{Tzitzeica, Dodd_Bullough_1977,
Zhiber_Shabat_1979, Mikhailov_1979}
\begin{equation}\label{Tz}
 H_{xy}=e^{H}-e^{-2H}
\end{equation}
is probably the most exotic integrable equation among those which can be
written in such a compact form. It looks deceptively similar to the Liouville
and the sine-Gordon equations, but the underlying spectral problem is of third
order and this makes all properties more complicated. In particular, the
problem of integrable discretization is rather difficult. The most natural
pattern for a discrete version is given obviously by quad-equations, that is,
difference equations on the square grid
\[
 Q(h,h_1,h_2,h_{12})=0
\]
where $h=h(n_1,n_2)$ and the subscript $k$ denotes the shift $n_k\to n_k+1$.
However, up to the author's knowledge, no such discretization was known till
now. The discretization proposed by Schief and Bobenko \cite{Schief_1999,
Bobenko_Schief_1999a, Bobenko_Schief_1999b} is very natural from the geometric
point of view, but it is given by a three-component system rather than one
scalar equation:
\begin{gather}
\nonumber
  hh_{12}(h_1h_2-h_1-h_2)+h_{12}+h-1-\frac{AB}{h}h_1h_2h_{12}=0,\\
\label{BS-Tz}
  \frac{A_2}{A}=\frac{h_1}{h},\quad \frac{B_1}{B}=\frac{h_2}{h}.
\end{gather}
The discretization through the permutability properties of the B\"acklund
transformations \cite{Boldin_Safin_Sharipov, Schief_1996a, Schief_1996b} also
leads to equations with several components, because the Tzitzeica equation does
not admit B\"acklund transformations of first order (in contrast to the
sine-Gordon equation).

On the other hand, the discrete theory is usually in perfect parallel with the
continuous one and therefore one should expect the existence of some
quad-equation analogous to (\ref{Tz}). The aim of this paper is to study the
following candidate for this role:
\begin{equation}\label{d-Tz}
 hh_{12}(c^{-1}h_1h_2-h_1-h_2)+h_{12}+h-c=0
\end{equation}
or, in the exponential form,
\[
 e^{H_1+H_2-C}-e^{H_1}-e^{H_2}=e^{C-H-H_{12}}-e^{-H}-e^{-H_{12}}.
\]
The parameter $c\ne0,\infty$ is essential and cannot be scaled, but the values
$c$, $-c$ are equivalent under the change of sign $h\to-h$ and $c$, $c^{-1}$
are equivalent under the reflection $h(n_1,n_2)\to 1/h(-n_1,n_2)$. The fixed
points $c=\pm1$ are special: at these points the equation degenerates into
\begin{equation}\label{d-L}
 hh_{12}(h_1-1)(h_2-1)=(h-1)(h_{12}-1)
\end{equation}
which is one of the form of the well known discrete Liouville equation. This is
a linearizable equation, namely, the substitution
$h=\dfrac{\tau_1\tau_2}{\tau\tau_{12}}$ maps solutions of the linear wave
equation $\tau_{12}-\tau_1-\tau_2+\tau=0$ into solutions of (\ref{d-L}). The
general solution of the latter is given therefore explicitly by the cross-ratio
\[
 h=\frac{(a_1-b)(a-b_2)}{(a-b)(a_1-b_2)}
\]
where $a=a(n_1)$ and $b=b(n_2)$ are arbitrary functions of one discrete
variable. At $c\ne\pm1$, no such explicit formula exists, however we will see
in Section \ref{s:bilinear} that the above substitution still makes sense and
brings the equation to the trilinear/bilinear forms.

The continuous limit is given by the following substitutions:
\[
 c\mapsto 1+\alpha\varepsilon^6,\quad
 h(n_1,n_2)\mapsto 1+\beta\varepsilon^2h(x,y),\quad
 x=\varepsilon n_1,\quad y=\varepsilon n_2.
\]
In the limit $\varepsilon\to0$, the terms up to $\varepsilon^5$ in equation
(\ref{d-Tz}) vanish identically and the coefficients at $\varepsilon^6$ recover
the equation
\[
 \beta^2(hh_{xy}-h_xh_y)=2\beta^3h^3-2\alpha
\]
which is the Tzitzeica equation (\ref{Tz}) up to the change $h=e^H$ and obvious
scalings (certainly, if $c\equiv1$ then $\alpha=0$ and we obtain the Liouville
equation, as it should be).

In the next section we discuss the zero curvature representation for equation
(\ref{d-Tz}). An associated differential-difference flow is derived in Section
\ref{s:dSK}. It serves as a Volterra lattice type discretization of the
Sawada--Kotera equation \cite{Sawada_Kotera} and its modification given in
Section \ref{s:M} defines a continuous symmetry of the discrete Tzitzeica
equation. The role of such symmetries in the theory of quad-equations is well
known, see e.g. \cite{LPSY}.

%-------------------------------------------------------------------------------
\section{Linear problems}\label{s:psi}

Integrability of equation (\ref{d-Tz}) is based on the zero curvature
representation which can be conveniently written as a system of second order
difference equations.

\begin{statement}
The discrete Tzitzeica equation (\ref{d-Tz}) is equivalent to the consistency
conditions of the following equations:
\begin{align}
\nonumber
 \psi_{11}-\mu\psi_1 &= \frac{h_1-c}{h_1(h-c)}(\psi_1-\mu\psi)
                       +\frac{c-\mu}{h-c}(\psi_{12}-\nu\psi_1),\\
\nonumber
 \psi_{12}+\psi ~&= h(\psi_1+\psi_2),\\
\label{psi12}
 \psi_{22}-\nu\psi_2 &= \frac{h_2-c}{h_2(h-c)}(\psi_2-\nu\psi)
                       +\frac{c-\nu}{h-c}(\psi_{12}-\mu\psi_2)
\end{align}
where
\[
 \mu=c-(c+1)\lambda,\quad \nu=c-(c-1)\lambda^{-1}
\]
and $\lambda$ is the spectral parameter.
\end{statement}

The proof is obtained by straightforward computation. Clearly, one can
substitute $\psi_{12}$ from the second equation into the other two and then the
system can be rewritten in the matrix form
\[
 \Psi_1=L\Psi,\quad \Psi_2=M\Psi
\]
where $\Psi=(\psi,\psi_1,\psi_2)^T$. The consistency condition then takes the
standard form of the discrete zero curvature representation $L_2M=M_1L$ with
$3\times3$ matrices. This equation is exactly equivalent to (\ref{d-Tz}).

The spectral problem for the Tzitzeica equation (\ref{Tz})
\begin{equation}\label{psixy}
 \psi_{xx}= v_x\psi_x+\lambda e^{-v}\psi_y,\quad
 \psi_{xy}= e^v\psi,\quad
 \psi_{yy}= \lambda^{-1}e^{-v}\psi_x+v_y\psi_y
\end{equation}
can be easily recovered by the continuous limit. In the matrix form we obtain
$\Psi_x=U\Psi$, $\Psi_y=V\Psi$ where $\Psi=(\psi,\psi_x,\psi_y)^T$, and the
zero curvature representation $U_y-V_x=[V,U]$ as the consistency condition.

Recall that equations (\ref{psixy}) are geometrically nothing but the Gauss
equations for the indefinite affine spheres $\psi:\mathbb R^2\to\mathbb R^3$
(the definite ones correspond to the elliptic version of the Tzitzeica
equation), see e.g. \cite{Rogers_Schief}.

In order to clarify the geometric meaning of system (\ref{psi12}), we compare
it with the Schief--Bobenko discretization of Gauss equations
\cite{Schief_1999, Bobenko_Schief_1999a, Bobenko_Schief_1999b}:
\begin{align}
\nonumber
 \psi_{11}-\psi_1 &= \frac{h_1-1}{h_1(h-1)}(\psi_1-\psi)
                    +\frac{\lambda A}{h-1}(\psi_{12}-\psi_1),\\
\nonumber
 \psi_{12}+\psi  ~&= h(\psi_1+\psi_2),\\
\label{BS-psi12}
 \psi_{22}-\psi_2 &= \frac{h_2-1}{h_2(h-1)}(\psi_2-\psi)
                    +\frac{\lambda^{-1}B}{h-1}(\psi_{12}-\psi_2).
\end{align}
In this case the consistency conditions give exactly system (\ref{BS-Tz}).
Equations (\ref{BS-psi12}) describe the so-called {\em discrete indefinite
affine spheres}, a class of discrete surfaces $\psi:{\mathbb Z}^2\to{\mathbb
R}^3$ characterized by two properties:\medskip

1) $\psi$ is a {\em discrete asymptotic net}, that is, the points $\psi$,
$\psi_1$, $\psi_{\overline1}$, $\psi_2$, $\psi_{\overline2}$ are coplanar for
any $(n_1,n_2)\in{\mathbb Z}^2$ (here, $\bar k$ denotes the backward shift
$n_k\to n_k-1$);\medskip

2) $\psi$ is a {\em discrete affine Lorentz harmonic net}, that is, all
discrete affine normals (defined as vectors $\psi_{12}-\psi_1-\psi_2+\psi$
attached to the centers of plaquettes
$\frac{1}{4}(\psi_{12}+\psi_1+\psi_2+\psi)$) meet in one point, the origin.
\medskip

Property 2) is expressed analytically by second equation (\ref{BS-psi12}) and
we see that it is satisfied for system (\ref{psi12}) as well.

Property 1) is expressed by first and last equations (\ref{BS-psi12}), with the
coefficients specified by use of the consistency condition. We see that this
property does not hold for the surface $\psi(n_1,n_2)$ defined by system
(\ref{psi12}), but it holds for the surface obtained by the gauge
transformation
\[
 \widetilde\psi(n_1,n_2)=\mu^{-n_1}\nu^{-n_2}\psi(n_1,n_2).
\]
Certainly, it is possible to rewrite the system in terms of $\widetilde\psi$,
then property 1) will be satisfied, but property 2) will be distorted.

Therefore, one can say that system (\ref{psi12}) defines a class of discrete
surfaces which are, up to a simple coordinate-dependent scaling, both discrete
asymptotic nets and discrete affine Lorentz harmonic nets, however in contrast
to the Schief--Bobenko case these properties are not fulfilled simultaneously.

To finish this section, we notice that elimination of $\psi_2$ and $\psi_{12}$
from two first equations (\ref{psi12}) brings to the following ordinary
difference equation of third order with respect to the shift $T_1$:
\[
 u\psi_{111}+\psi_{11}=\mu(\psi_1+u\psi),\quad
 u:=\frac{h_{11}(c-h_1)}{h_{11}h_1h-c}
\]
(certainly, an analogous equation holds for the shift $T_2$). This linear
problem is considered in more details in Section \ref{s:dSK}.

%-------------------------------------------------------------------------------
\section{Bilinear equations}\label{s:bilinear}

Consider the continuous case first. It is well known (see e.g.
\cite{Bobenko_Schief_1999a,Hirota_Takahashi_2005}) that the substitution
$h=-2(\log\tau)_{xy}$ maps the algebraic form of Tzitzeica equation
\[
 hh_{xy}-h_xh_y=h^3-1
\]
into the trilinear form
\begin{equation}\label{Tz3}
 4\det\begin{pmatrix}
  \tau_{yy} & \tau_{xyy} & \tau_{xxyy} \\
  \tau_y    & \tau_{xy}  & \tau_{xxy}  \\
  \tau      & \tau_x     & \tau_{xx}
 \end{pmatrix}=\tau^3.
\end{equation}
The fact which apparently has not been paid attention before is that a couple
of simpler bilinear equations can be added,
\begin{align}
\nonumber
 3(\tau_{xy}\tau_{xx}-\tau_x\tau_{xxy})&=\tau_y\tau_{xxx}-\tau\tau_{xxxy},\\
\label{Tz2}
 3(\tau_{xy}\tau_{yy}-\tau_y\tau_{xyy})&=\tau_x\tau_{yyy}-\tau\tau_{xyyy}
\end{align}
which are consistent with (\ref{Tz3}). These follow from the conservation laws
\[
 \Bigl(\frac{h_{xx}}{h}\Bigr)_y=3D_x(h),\quad
 \Bigl(\frac{h_{yy}}{h}\Bigr)_x=3D_y(h)
\]
which admit integration after the substitution:
\[
 \frac{h_{xx}}{h}=-6(\log\tau)_{xx}+a(x),\quad
 \frac{h_{yy}}{h}=-6(\log\tau)_{yy}+b(y).
\]
Notice that $\tau$-function is defined in (\ref{Tz3}) up to the multiplication
by arbitrary functions on $x$ and on $y$, and this freedom can be fixed by
setting $a=b=0$ without loss of generality. Now substituting $\tau$ again
yields relations
\[
 (\log\tau)_{xxxy}=-6(\log\tau)_{xx}(\log\tau)_{xy},\quad
 (\log\tau)_{xyyy}=-6(\log\tau)_{yy}(\log\tau)_{xy}
\]
which are (\ref{Tz2}).

In the discrete case the picture is very similar. The substitution
$h=\dfrac{\tau_1\tau_2}{\tau\tau_{12}}$ brings (\ref{d-Tz})
\[
 hh_{12}(c^{-1}h_1h_2-h_1-h_2)+h_{12}+h-c=0
\]
to the trilinear form
\[
 c^{-1}\tau_{22}\tau_{12}\tau_{11}+\tau\tau_{122}\tau_{112}+\tau_1\tau_2\tau_{1122}
 =\tau_{11}\tau_2\tau_{122}+\tau_1\tau_{22}\tau_{112}+c\tau\tau_{12}\tau_{1122}
\]
which can be rewritten as
\begin{equation}\label{dTz3}
 \det\begin{pmatrix}
  \tau_{22} & \tau_{122} & \tau_{1122} \\
  \tau_2    & c^{-1}\tau_{12}  & \tau_{112} \\
  \tau      & \tau_1     & \tau_{11}
 \end{pmatrix}=(c-c^{-1})\tau\tau_{12}\tau_{1122}.
\end{equation}
The additional bilinear equations
\begin{align}
\nonumber
 \tau_{11}\tau_{12}-c\tau_1\tau_{112}& =c\tau\tau_{1112}-\tau_{111}\tau_2,\\
\label{dTz2}
 \tau_{12}\tau_{22}-c\tau_2\tau_{122}& =c\tau\tau_{1222}-\tau_{222}\tau_1
\end{align}
can be extracted from the multiplicative conservation laws
\begin{equation}\label{uuhh}
 \frac{u_2}{u}=\frac{h}{h_{11}},\quad u=\frac{h_{11}(c-h_1)}{h_{11}h_1h-c}\,;\qquad
 \frac{v_1}{v}=\frac{h}{h_{22}},\quad v=\frac{h_{22}(c-h_2)}{h_{22}h_2h-c}\,.
\end{equation}
These admit integration after the substitution:
\[
 u=a(n_1)\frac{\tau\tau_{111}}{\tau_{11}\tau_1},\quad
 v=b(n_2)\frac{\tau\tau_{222}}{\tau_{22}\tau_2},
\]
and again, since the $\tau$-function is defined up to multiplication by
arbitrary functions on $n_1$ and $n_2$, hence one can chose integration
constants $a=b=1$ without loss of generality. Replacing $h$ in the expressions
for $u,v$ yields (\ref{dTz2}).

In the case of Bobenko--Schief discretization (\ref{BS-Tz}) the substitution
for $h$ is supplemented by $A=a\dfrac{\tau^2_1}{\tau\tau_{11}}$,
$B=b\dfrac{\tau^2_2}{\tau\tau_{22}}$ with arbitrary $a(n_1)$, $b(n_2)$, and
this leads to the trilinear equation \cite{Schief_1999, Bobenko_Schief_1999a,
Bobenko_Schief_1999b}
\begin{equation}\label{BS-dTz3}
 \det\begin{pmatrix}
  \tau_{22} & \tau_{122} & \tau_{1122} \\
  \tau_2    & \tau_{12}  & \tau_{112} \\
  \tau      & \tau_1     & \tau_{11}
 \end{pmatrix}=ab\tau^3_{12}.
\end{equation}
The bilinear equations are not known for this discretization. Clearly, both
equations (\ref{dTz3}) and (\ref{BS-dTz3}) go to (\ref{Tz3}) under the
corresponding continuous limits.

%-------------------------------------------------------------------------------
\section{A difference analog of Sawada--Kotera equation}\label{s:dSK}

In the rest of the paper we change the notation: now we will consider only one
discrete variable $n$ (identified, say, with $n_1$) and it is more convenient
to reserve subscripts for the order of shift along this variable rather than to
distinguish between one-step shifts along different variables $n_k$ as before.
For instance, new $h_4$ is the same as old $h_{1111}$ and $u_{-3}$ is the same
as $u_{\overline{111}}$.

The goal of this section is to derive the differential-difference equation
\begin{equation}\label{d-SK}
 u_t=u^2(u_2u_1-u_{-1}u_{-2})-u(u_1-u_{-1}).
\end{equation}
Its close relation to the discrete Tzitzeica equation will be revealed in the
next section, however, this lattice certainly deserves study on its own. It can
be interpreted as a difference analog of the Sawada--Kotera equation
\cite{Sawada_Kotera}
\[
 U_\tau=U_{xxxxx}+5UU_{xxx}+5U_xU_{xx}+5U^2U_x
\]
which appears under the following continuous limit at $\varepsilon\to0$:
\[
 u(n,t)=\frac{1}{3}+\frac{\varepsilon^2}{9}U\Bigl(x-\frac{4}{9}\varepsilon t,
    \tau+\frac{2\varepsilon^5}{135}t\Bigr),\quad x=\varepsilon n.
\]
Recall that both flows
\[
 u_{t'}=u(u_1-u_{-1})\qquad \text{and}\qquad u_{t''}=u^2(u_2u_1-u_{-1}u_{-2})
\]
are very well known integrable equations: the first one is the celebrated
Volterra lattice and the second one is the modified Narita--Bogoyavlensky
lattice (see \cite{Yamilov_1983, Yamilov_2006, Suris_2003} for a detailed
theory and the bibliography). In both cases, the continuous limit is the
Korteweg--de Vries equation $U_t=U_{xxx}+6UU_x$. However, these lattices are
not members of one and the same hierarchy, that is the flows $\partial_{t'}$
and $\partial_{t''}$ do not commute. Therefore, one should not expect that
their linear combination (\ref{d-SK}) is integrable. But it is, and we will
prove it by constructing the Lax representation.

The starting point is the difference spectral problem
\begin{equation}\label{psi3}
 u\psi_3+\psi_2=\mu(\psi_1+u\psi).
\end{equation}
This is a special reduction of a general third order problem, in parallel with
the continuous case of the Sawada--Kotera equation. We write it in the operator
form
\[
 P\psi=\mu Q\psi,\quad P=(uT+1)T^2,\quad Q=T+u
\]
where $T:n\to n+1$ is the shift operator. The isospectral deformations are
defined by the Lax equations
\[
 D_t(Q^{-1}P)=[A,Q^{-1}P] \quad \Leftrightarrow\quad
 P_t=BP-PA,\quad Q_t=BQ-QA.
\]
These equations are equivalent, for the above $P,Q$, to the system
\[
 u_t=B(T+u)-(T+u)A,\quad B(T^2-1)=(T+u)AT-(uT+1)A_2.
\]
We assume that both $A$ and $B$ are difference operators, that is, Laurent
polynomials in $T$. The notation like $A_2$ is used for the operators with
shifted coefficients, so that $T^2A=A_2T^2$. The second equation can be solved
as follows:
\[
 A=F(T-T^{-1}),\quad B=F_1T+u(F-F_3)-F_2T^{-1},
\]
and then the substitution into the first equation yields
\begin{equation}\label{ut}
 u_t=TFu+uFT^{-1}-uF_3T-T^{-1}F_3u+F_1-F_2+u(F-F_3)u.
\end{equation}
Notice that this equation admits the reduction $F^*=F$, with respect to the
usual conjugation of difference operators: $T^*=T^{-1}$, $(FG)^*=G^*F^*$ and
$f^*=f$ for functions $f$. Let operator $F$ be of the form
\[
 F=f^{(k,k)}T^k+\dots+f^{(k,1)}T+f^{(k,0)}+f^{(k,1)}_{-1}T^{-1}
  +\dots+f^{(k,k)}_{-k}T^{-k}, \quad k>0.
\]
Then collecting coefficients at $T^{k+1},\dots,T$ in equation (\ref{ut}) brings
to the recurrent relations
\begin{align*}
 u_{k+1}f^{(k,k)}_1-uf^{(k,k)}_3 &= 0,\\
 u_kf^{(k,k-1)}_1-uf^{(k,k-1)}_3 &= f^{(k,k)}_2-f^{(k,k)}_1+uu_k(f^{(k,k)}_3-f^{(k,k)}),\\
 u_jf^{(k,j-1)}_1-uf^{(k,j-1)}_3 &= f^{(k,j)}_2-f^{(k,j)}_1+uu_j(f^{(k,j)}_3-f^{(k,j)})\\
   &\qquad +u_jf^{(k,j+1)}_2-uf^{(k,j+1)},\qquad j=1,\dots,k-1
\end{align*}
and vanishing of the free term yields the desired equation for $u$
\[
 u_{t_k}=2u(f^{(k,1)}-f^{(k,1)}_2)+u^2(f^{(k,0)}-f^{(k,0)}_3)+f^{(k,0)}_1-f^{(k,0)}_2.
\]
We are interested only in a local evolution, that is, the recurrent relations
must define all $f^{(k,j)}$ as functions of a finite set of $u_i$. It is easy
to see that the first equation for $f^{(k,k)}$ can be solved if and only if $k$
is odd, namely
\[
 f^{(k,k)}=u_{-1}\cdots u_{k-4}u_{k-2},\quad k=2m+1>0.
\]
In the simplest case $k=1$ we find
\[
 f^{(1,1)}=u_{-1},\quad f^{(1,0)}=1-u_{-2}u_{-1}
\]
and the corresponding flow is exactly (\ref{d-SK}). So, we arrive to the
following statement.

\begin{statement}
Lattice (\ref{d-SK}) governs the isospectral deformation of the linear problem
(\ref{psi3}) defined by the equation
\begin{equation}\label{psit}
 \psi_t=A(\psi)=(u_{-1}T+1-u_{-1}u_{-2}+u_{-2}T^{-1})(T-T^{-1})(\psi).
\end{equation}
\end{statement}

The higher flows can be computed analogously. At $k=3$ one finds
\begin{gather*}
 f^{(3,3)}=u_1u_{-1},\quad f^{(3,2)}=w+w_{-1},\quad
 f^{(3,1)}=1-u_{-1}(w+w_{-1}+w_{-2}),\\
 f^{(3,0)}=u_{-1}u_{-2}(w+w_{-1}+w_{-2}+w_{-3})
\end{gather*}
where $w=u(1-u_1u_{-1})$ and
\begin{gather*}
 u_{t_3}=u\bigl(w_1(w_3+w_2+w_1+w)-w_{-1}(w+w_{-1}+w_{-2}+w_{-3})\\
   -u_1(w_3+w_{-1})+u_{-1}(w_1+w_{-3})\bigr).
\end{gather*}
One can check directly that this is a commuting flow for (\ref{d-SK}) indeed.
It should be noted that its right hand side is a sum of three homogeneous
polynomials:
\[
 u_{t_3}=P^{(3)}+P^{(5)}+P^{(7)},\quad \mathop{\rm deg}P^{(m)}=m,
\]
where $P^{(3)}$ and $P^{(5)}$ define respectively the symmetries for the
Volterra and the Narita--Bogoyavlensky lattices
\begin{gather*}
 u_{t'_3}=P^{(3)}=u(u_1(u_2+u_1+u)-u_{-1}(u+u_{-1}+u_{-2})),\\
 u_{t''_3}=P^{(7)}=u^2u_1u^2_2u_3u_4+\dots
\end{gather*}

%-------------------------------------------------------------------------------
\section{Modified lattice}\label{s:M}

The role of discrete Miura type transformations for lattice (\ref{d-SK}) is
played by the substitutions
\[
 M^-:\quad u=\frac{h_2(c-h_1)}{h_2h_1h-c},\qquad
 M^+:\quad \hat u=\frac{(c-h_1)h}{h_2h_1h-c}
\]
(the first one already appeared in the end of Section \ref{s:psi}).

\begin{statement}
The substitutions $M^\pm$ map solutions of the lattice
\begin{equation}\label{ht}
 h_t=\frac{h(c-h)}{h_1hh_{-1}-c}\left(
  \frac{h(c-h_1)(c-h_{-1})(h_2h_1-h_{-1}h_{-2})}{(h_2h_1h-c)(hh_{-1}h_{-2}-c)}
  -h_1+h_{-1}\right)
\end{equation}
into solutions of lattice (\ref{d-SK}).
\end{statement}
\begin{proof}
One can easily check that if $\phi$ is a particular solution of linear problem
(\ref{psi3}) corresponding to the value of spectral parameter $\mu=1/c$ then
the quotient $h=\phi/\phi_1$ is related to the potential $u$ by the
substitution $M^-$. Therefore, the time evolution of $h$ can be found by use of
equation (\ref{psit}) for $\phi_t$ which guarantees that $h$ satisfies some
modified lattice equation in the form of a conservation law
\begin{equation}\label{htS}
 (\log h)_t=(T-1)S(h_1,h,h_{-1},h_{-2})
\end{equation}
where
\begin{align*}
 S&= -\frac{\phi_t}{\phi}
   = \frac{1}{\phi}(u_{-1}T+1-u_{-1}u_{-2}+u_{-2}T^{-1})(\phi_{-1}-\phi_1)\\
  &= u_{-1}\Bigl(1-\frac{1}{h_1h}\Bigr)+u_{-2}(h_{-1}h_{-2}-1)
    +(1-u_{-1}u_{-2})\Bigl(h_{-1}-\frac{1}{h}\Bigr).
\end{align*}
Replacing $u$ and some algebra bring to equation (\ref{ht}).

The second substitution follows either from the involution $h\to h^{-1}$, $c\to
c^{-1}$ which preserves equation (\ref{ht}) or from the form invariance of
linear problem (\ref{psi3}) with respect to the change $n\to-n$.
\end{proof}

In order to relate lattice (\ref{ht}) with the discrete Tzitzeica equation
(\ref{d-Tz}) we notice that $\hat u/u=h/h_2$ and compare this with the first of
conservation laws (\ref{uuhh}). It is clear that if the discrete variable $n$
in (\ref{ht}) is identified with $n_1$ in (\ref{d-Tz}) then the map $u\to\hat
u$ can be identified with the shift $T_2$ along the second discrete variable
$n_2$. This argument is not quite rigorous, since the conservation law is not
exactly equivalent to equation (\ref{d-Tz}). Nevertheless, the following
statement holds true.

\begin{statement}
The discrete Tzitzeica equation (\ref{d-Tz}) admits the evolution symmetry
(\ref{ht}) along any of the coordinate directions $n=n_1$ or $n=n_2$, that is,
the differentiation $D_t$ in virtue of lattice (\ref{ht}) is consistent with
the discrete equation:
\[
 D_t(Q)|_{Q=0}=0.
\]
\end{statement}

The proof of this identity is straightforward, although rather involved:
$D_t(Q)$ contains variables $h$ in 12 points $(n_1+k,n_2)$, $(n_1+k,n_2+1)$,
$k=-2,\dots,3$, so that 5 copies of the quad-equation
$T^{-2}(Q)=\dots=T^2(Q)=0$ are used.

It is worth noticing that an infinite sequence of conservation laws for lattice
(\ref{d-SK}) can be extracted from the single conservation law (\ref{htS}) for
the modified lattice, by means of the classical trick with the inversion of
Miura transformation as a power series in the spectral parameter
\cite{Miura_Gardner_Kruskal}. Indeed, the equation $u=M^-(h)$ can be solved
with respect to $h$ as the formal power series
\[
 h=-\frac{1}{u}(1+ch^{(1)}+c^2h^{(2)}+\dots)
\]
with the coefficients recursively found as polynomials in $u_j$:
\begin{align*}
 h^{(1)}&= u_1(1-uu_2),\\
 h^{(2)}&= -u_1u_2\bigl((1-u_2u)(1-u_3u_1)-u_3u(1-u_4u_2)\bigr),~\dots
\end{align*}
Then the coefficients of the expansion
\[
 \log h=-\log(-u)+ch^{(1)}+c^2\Bigl(h^{(2)}-\frac{1}{2}(h^{(1)})^2\Bigr)
 +\dots
\]
provide densities of the conservation laws $D_t(\rho^{(k)})=(T-1)\sigma^{(k)}$:
\[
 \rho^{(0)}= \log u,\quad \rho^{(1)}= u(1-u_1u_{-1}),~\dots
\]

%-------------------------------------------------------------------------------
\section{Concluding remarks}

It should be noted that the discrete Tzitzeica equation (\ref{d-Tz}) falls out
of the classification scheme based on the notion of 3D-consistency. Indeed,
this notion is essentially equivalent to the existence of a zero curvature
representation with $2\times2$ matrices \cite{Nijhoff_Walker}, while the
construction in Section \ref{s:psi} brings to the $3\times3$ matrices and it is
difficult to expect that the size can be reduced. Up to the author's knowledge,
this is the first example of a quad-equation associated with a third-order
spectral problem. According to \cite{ABS_2009}, any `generic' quad-equation can
be included into a consistent triple, but the situation remains unclear for
equations with degenerate biquadratics. In this terminology, equation
(\ref{d-Tz}) is degenerate, and this example demonstrates that this class of
equations can be more complicated than it seems at a first glance. The
alternative approaches to the integrability of quad-equations are developed,
e.g. in \cite{Levi_Yamilov, Mikhailov_Wang_Xenitidis, Habibullin_Gudkova} and
one may hope that more examples of such kind will be discovered by these
methods.

An open problem is to construct the B\"acklund transformations for equation
(\ref{d-Tz}). Probably, the commutativity of these transformations can be
formulated as 3D-consistency of equation (\ref{d-Tz}) with some discrete
equations of second order in shifts rather than quad-equations.

The exhaustive classification of integrable Volterra-type lattices
$u_t=f(u_1,u,u_{-1})$ was obtained by Yamilov \cite{Yamilov_1983,
Yamilov_2006}. However, a little is known about the higher order lattices, even
in the polynomial case, so that the study of lattices like (\ref{d-SK}) is of
interest as well.

\section*{Acknowledgements}

This work was supported by the grant NSh-6501.2010.2.

\end{document}